\documentclass[runningheads]{llncs}

\usepackage[ruled]{algorithm}
\usepackage{algpseudocode}
\usepackage[T1]{fontenc}
\usepackage{amsmath} 
\usepackage{amssymb}  
\usepackage{hyperref}

\hypersetup{
    colorlinks=true,
    linkcolor=blue,
    filecolor=magenta,      
    urlcolor=cyan,
    pdftitle={Overleaf Example},
    pdfpagemode=FullScreen,
    }
\usepackage{siunitx}
\usepackage{threeparttable}
\usepackage{multirow}
\usepackage{makecell}
\usepackage{booktabs}
\usepackage{array}
\newcolumntype{C}[1]{>{\centering\let\newline\\\arraybackslash\hspace{0pt}}m{#1}}
\newcolumntype{L}[1]{>{\raggedright\let\newline\\\arraybackslash\hspace{0pt}}m{#1}}

\usepackage{graphicx}
\begin{document}
\title{
MI-SegNet: Mutual Information-Based US Segmentation for Unseen Domain Generalization
}
%
\titlerunning{MI-SegNet: Mutual Information-Based US Segmentation}
%
\author{Anonymous}
\author{Yuan Bi\inst{1} \and 
Zhongliang Jiang\thanks{Corresponding Author}\inst{1} \and 
Ricarda Clarenbach\inst{2} \and 
Reza Ghotbi\inst{2} \and 
Angelos Karlas\inst{3} \and 
Nassir Navab\inst{1}} 
\authorrunning{Y. Bi et al.}
%
\institute{Chair for Computer-Aided Medical Procedures and Augmented Reality, Technical University of Munich, Munich, Germany \and
Clinic for Vascular Surgery, Helios Klinikum M{\"u}nchen West, Munich, Germany \and
Department for Vascular and Endovascular Surgery, rechts der Isar University Hospital, Technical University of Munich, Munich, Germany}
\maketitle              
\begin{abstract}
Generalization capabilities of learning-based medical image segmentation across domains are currently limited by the performance degradation caused by the domain shift, particularly for ultrasound (US) imaging. The quality of US images heavily relies on carefully tuned acoustic parameters, which vary across sonographers, machines, and settings. To improve the generalizability on US images across domains, we propose MI-SegNet, a novel mutual information (MI) based framework to explicitly disentangle the anatomical and domain feature representations; therefore, robust domain-independent segmentation can be expected. Two encoders are employed to extract the relevant features for the disentanglement. The segmentation only uses the anatomical feature map for its prediction. In order to force the encoders to learn meaningful feature representations a cross-reconstruction method is used during training. Transformations, specific to either domain or anatomy are applied to guide the encoders in their respective feature extraction task. Additionally, any MI present in both feature maps is punished to further promote separate feature spaces.
We validate the generalizability of the proposed domain-independent segmentation approach on several datasets with varying parameters and machines.
Furthermore, we demonstrate the effectiveness of the proposed MI-SegNet serving as a pre-trained model by comparing it with state-of-the-art networks.\footnote{The code is available at: \url{https://github.com/yuan-12138/MI-SegNet}}

\end{abstract}
\section{Introduction}
Deep neural networks (DNNs) have achieved phenomenal success in image analysis and comparable human performance in many semantic segmentation tasks. However, based on the assumption of DNNs, the training and testing data of the network should come from the same probability distribution~\cite{valiant1984theory}. The generalization ability of DNNs on unseen domains is limited. The lack of generalizability hinders the further implementation of DNNs in real-world scenarios. 

\par
Ultrasound (US), as one of the most popular means of medical imaging, is widely used in daily medical practice to diagnose internal organs, such as vascular structures. Compared to other imaging methods, e.g., computed tomography (CT) and magnetic resonance imaging (MRI), US shows its advantages in terms of being radiation-free and portable. To accurately and robustly extract the vascular lumen for diagnosis, the Doppler signal~\cite{jiang2023dopus} and artery pulsation signal~\cite{huang2023motion} were employed to facilitate vessel segmentation. However, the US image quality is operator-dependent and sensitive to inter-machine and inter-patient variations. Therefore, the performance of the US segmentation is often decayed due to the domain shift caused by the inconsistency between the training and test data~\cite{huang2022online}.


\par
\subsubsection{Data Augmentation} One of the most common ways of improving the generalization ability of DNNs is to increase the variability of the dataset~\cite{zhang2021understanding}. However, in most clinical cases, the number of data is limited. Therefore, data augmentation is often used as a feasible method to increase diversity. Zhang~\emph{et al.} proposed BigAug~\cite{zhang2020generalizing}, a deep stacked transformation method for 3D medical image augmentation. By applying a wide variety of augmentation methods to the single source training data, they showed the trained network is able to increase its performance on unseen domains. In order to take the physics of US into consideration, Tirindelli~\emph{et al.} proposed a physics-inspired augmentation method to generate realistic US images~\cite{tirindelli2021rethinking}. 

\subsubsection{Image-Level Domain Adaptation} To make the network generalizable to target domains that are different from the source domain, the most intuitive way is to transfer the image style to the same domain. The work from Chen~\emph{et al.} achieved impressive segmentation results in MRI to CT adaptation by applying both image and feature level alignment~\cite{chen2020unsupervised}. To increase the robustness of segmentation networks for US images, Yang~\emph{et al.} utilized a rendering network to unify the image styles of training and test data so that the model is able to perform equally well on different domains~\cite{yang2018generalizing}. Velikova~\emph{et al.} extended this idea by defining a common anatomical CT-US space so that the labeled CT data can be exploited to train a segmentation network for US images~\cite{velikova2022cactuss}.

\subsubsection{Feature Disentanglement} Instead of solving the domain adaptation problem directly at the image-level, many researchers focused on disentangling the features in latent space, forcing the network to learn the shared statistical shape model across different domains~\cite{bengio2013representation}. One way of realizing this is through adversarial learning~\cite{huang2018multimodal}\cite{lee2018diverse}\cite{ning2021new}\cite{zhao2022multi}.
However, adversarial learning optimization remains difficult and unstable in practice~\cite{lezama2018overcoming}. A promising solution for decoupling latent representations is to minimize a metric that can explicitly measure the shared information between different features. 
Mutual information (MI), which measures the amount of shared information between two random variables~\cite{kraskov2004estimating}, suits this demand. 
Previous researches have exploited its usage in increasing the generalizability for classification networks when solving the vision recognition~\cite{cha2022domain}\cite{liu2021mutual}\cite{peng2019domain} and US image classification~\cite{meng2020mutual} problems.
In this study, we investigate the effective way to integrate MI into a segmentation network in order to improve the adaptiveness on unseen images. 

\par
To solve the performance drop caused by the domain shift in segmentation networks, the aforementioned methods require a known target domain, e.g., CT~\cite{chen2020unsupervised}, MRI~\cite{ning2021new}, contrast enhanced US~\cite{zhao2022multi}. However, compared to MRI and CT, the image quality of US is more unstable and unpredictable. It is frequently observed that the performance of a segmentation network decreases dramatically for the US images acquired from a different machine or even with a different set of acquisition parameters. In such cases, it is impractical to define a so-called target US domain. 
Here we introduce MI-SegNet, an MI-based segmentation network, to address the domain shift problem in US image segmentation. Specifically, the proposed network extracts the disentangled domain (image style) and anatomical (shape) features from US images. The segmentation mask is generated based on the anatomical features, while the domain features are explicitly excluded. Thereby, the segmentation network is able to understand the statistical shape model of the target anatomy and generalize to different unseen scenarios. The ablation study shows that the proposed MI-SegNet is able to increase the generalization ability of the segmentation network in unseen domains.

\section{Method}

Our goal is to train a segmentation network that can generalize to unseen domains and serve as a good pre-trained model for downstream tasks, while the training dataset only contains images from a single domain. To this end, the training framework should be designed to focus on the shape of the segmentation target rather than the background or appearance of the images. Following this concept of design, we propose MI-SegNet. During the training phase, a parameterised data transformation procedure is undertaken for each training image ($x$). Two sets of parameters are generated for spatial ($a_1, a_2$) and domain ($d_1, d_2$) transformation respectively. For individual input, four transformed images ($x_{a_1d_1}, x_{a_2d_2}, x_{a_1d_2}, x_{a_2d_1}$) are created according to the four possible combinations of the spatial and domain configuration parameters. Two encoders ($E_a, E_d$) are applied to extract the anatomical features ($f_{a_1}, f_{a_2}$) and domain features ($f_{d_1}, f_{d_2}$) separately. The mutual information between the extracted anatomical features and the domain features from the same image is computed using mutual information neural estimator (MINE)~\cite{belghazi2018mutual} and minimized during training. Only the anatomical features are used to compute segmentation masks ($m_1, m_2$). The extracted anatomical and domain features are then combined and fed into the generator network ($G$) to reconstruct the images ($\widehat{x}_{a_1d_1}, \widehat{x}_{a_1d_2}, \widehat{x}_{a_2d_1}, \widehat{x}_{a_2d_2}$) accordingly. Since the images are transformed explicitly, it is possible to provide direct supervision to the reconstructed images. Notably, only two of the transformed images ($x_{a_1d_1}, x_{a_2d_2}$) are fed into the network, while the other two ($x_{a_1d_2}, x_{a_2d_1}$) are used as ground truth for reconstructions.
\begin{figure}
\includegraphics[width=\textwidth]{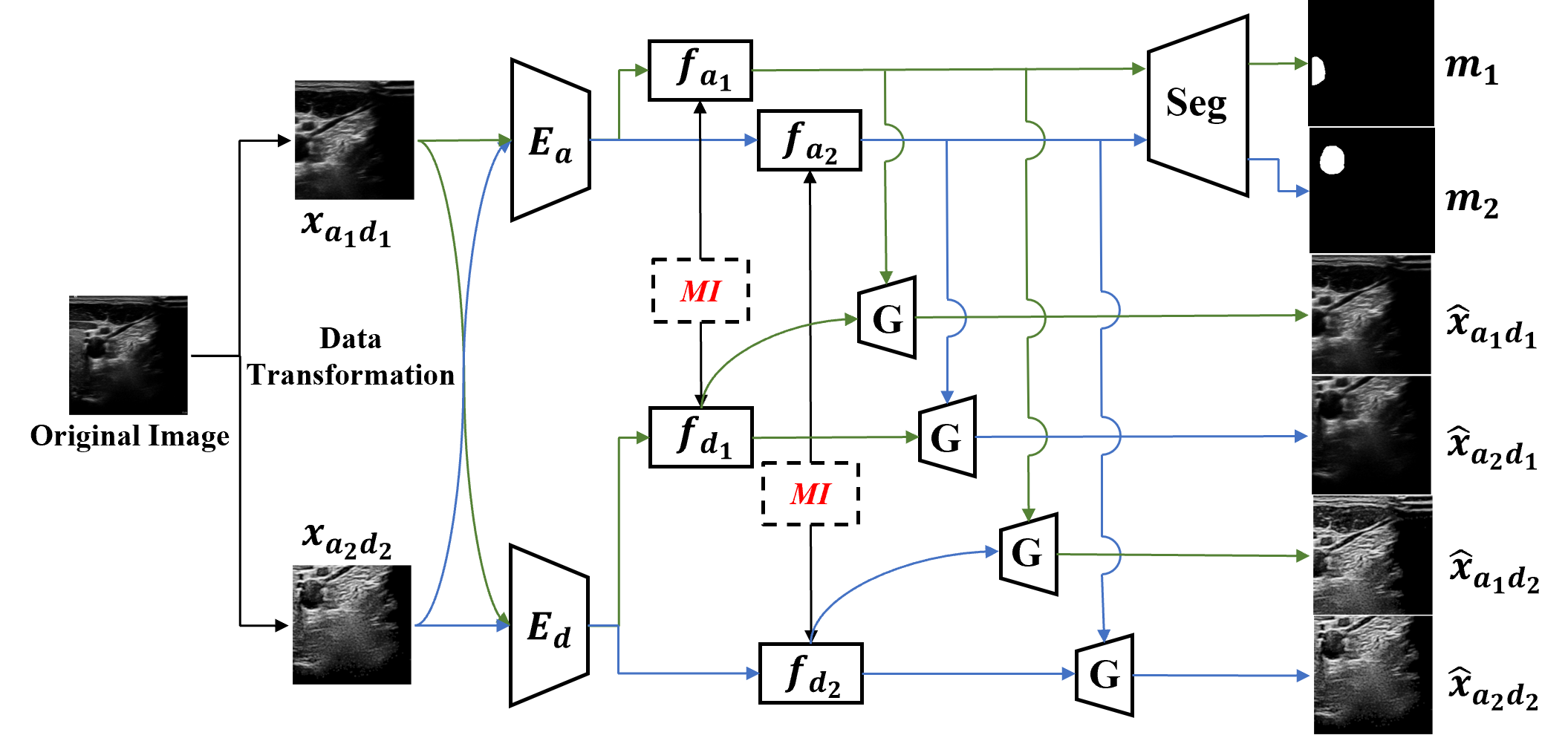}
\caption{
Network structure of MI-SegNet. The green and blue arrows represent the data flow of the first ($x_{a_1d_1}$) and the second input image ($x_{a_2d_2}$), respectively.
} \label{Fig:Network}
\end{figure}

\subsection{Mutual Information}
In order to decouple the anatomical and domain features intuitively, a metric that can evaluate the dependencies between two variables is needed. Mutual information, by definition, is a metric that measures the amount of information obtained from a random variable by observing another random variable. The MI is defined as the Kullback-Leibler (KL) divergence between the joint distribution and the product of marginal distributions of random variables $f_a$ and $f_d$:
\begin{equation}\label{Eq:MIDefinition}
\mathcal{MI}(f_a;f_d) = \mathcal{KL}(p(f_a,f_d)\|p(f_a)\otimes p(f_d))
\end{equation}
where $p(f_a,f_d)$ is the joint distribution and $p(f_a)\otimes p(f_d)$ is the product of the marginal distributions. Based on the Donsker-Varadhan representation~\cite{DonskerVaradhan1983}, the lower bound of MI can be represented as:
\begin{equation}\label{Eq:DonskerVaradhan}
\mathcal{MI}(f_a;f_d)  \geq E_{p(f_a,f_d)}[\mathcal{T}(f_a,f_d)]-\log(E_{p(f_a)\otimes p(f_d)}[e^{\mathcal{T}(f_a,f_d)}])
\end{equation}
where $\mathcal{T}$ is any arbitrary given continuous function. By replacing $\mathcal{T}$ with a neural network $\mathcal{T}_{\theta_{MINE}}$ and applying Monte Carlo method~\cite{peng2019domain}, the lower bound can be calculated as:
\begin{equation}\label{Eq:MonteCarlo}
\widehat{\mathcal{MI}(f_a;f_d)}=\frac{1}{N}\sum_{i=1}^N \mathcal{T}_{\theta_{MINE}}(f_a,f_d) - \log \frac{1}{N} \sum_{i=1}^N e^{\mathcal{T}_{\theta_{MINE}}(f_a',f_d')}
\end{equation}
where $(f_a,f_d)$ are drawn from the joint distribution and $(f_a',f_d')$ are drawn from the product of marginal distributions. By updating the parameters $\theta_{MINE}$ to maximize the lower bound expression in Eq.~\ref{Eq:MonteCarlo}, a loose estimation of MI is achieved, also known as MINE~\cite{belghazi2018mutual}. 

To force the anatomical and domain encoders to extract decoupled features, the MI is served as a loss to update the weights of these two encoder networks. The loss is defined as:
\begin{equation}\label{Eq:MILoss}
\mathcal{L}_{MI} = \widehat{\mathcal{MI}(f_a;f_d)}
\end{equation}

\subsection{Image Segmentation and Reconstruction}\label{section:ImageRec}
To make the segmentation network independent of the domain information, 
the domain features are excluded when generating the segmentation mask.
Here, the segmentation loss $\mathcal{L}_{seg}$ is defined in the combined form of dice loss $\mathcal{L}_{dice}$ and binary cross-entropy loss $\mathcal{L}_{bce}$.  
\begin{equation}\label{Eq:SegmentationLoss}
\begin{split}
\mathcal{L}_{seg} &= \mathcal{L}_{dice} + \mathcal{L}_{bce}\\
&= 1 - \frac{1}{N}\sum_{n=1}^N \frac{2 l_n m_n + s}{l_n + m_n +s} -\frac{1}{N}\sum_{n=1}^N (l_n \log m_n + (1-l_n)\log (1-m_n))
\end{split}
\end{equation}
where $l$ is the ground truth label, $m$ represents the predicted mask, $s$ is added to ensure the numerical stability, and $N$ is the mini batch size.

To ensure that the extracted anatomical and domain features can contain all the information of the input image, a generator network is used to reconstruct the image based on both features. The reconstruction loss is then defined as:
\begin{equation}\label{Eq:RecLoss}
\mathcal{L}_{rec} = \frac{1}{N}\sum_{n=1}^N \frac{1}{wh} (x_n-\widehat{x}_n)^2
\end{equation}
where $x_n$ is the ground truth image, $\widehat{x}_n$ is the reconstructed image, $w$ and $h$ are the width and height of the image in pixel accordingly.

\subsection{Data Transformation}
Since the training dataset only contains images from one single domain, it is necessary to enrich the diversity of the training data so that overfitting can be prevented and the generalization ability is increased. The transformation methods are divided into two categories, domain and spatial transformations. Each transformation ($T$) is controlled by two parameters, probability ($p$) and magnitude ($\lambda$).

\subsubsection{Domain Transformations} aim to transfer the single domain images to different domain styles. Five types of transformation methods are involved in this aspect, i.e., \textit{blurriness}, \textit{sharpness}, \textit{noise level}, \textit{brightness}, and \textit{contrast}. The implementations are identical to~\cite{zhang2020generalizing}, except the Gaussian noise is replaced by Rayleigh noise.
The possibility of all the domain transformations are empirically set to $10\%$.

\subsubsection{Spatial Transformations} mainly consist of two parts, \textit{crop} and \textit{flip}. For cropping, a window with configurable sizes ($[0.7,0.9]$ of the original image size) is randomly masked on the original image. Then the cropped area is resized to the original size to introduce varying shapes of anatomy. Here $\lambda$ controls the size and the position of the cropping window. Besides cropping, horizontal flipping is also involved. Unlike domain transformations, the labels are also transformed accordingly by the same spatial transformation. The probability ($p$) of flipping is $5\%$, while the $p$ for cropping is $50\%$ to introduce varying anatomy sizes.The images are then transformed in a stacked way:
\begin{equation}\label{Eq:DataAugmentation}
x_{aug} = T^{(P[n],\Lambda[n])}(T^{(P[n-1],\Lambda[n-1])}\cdots (T^{(P[1],\Lambda[1])}(x)))
\end{equation}
where $n=7$ represents the seven different transformation methods involved in our work, $\Lambda=[\lambda_n,\lambda_{n-1},\cdots, \lambda_1]$ represents the magnitude parameter, and $P=[p_n,p_{n-1},\cdots,p_1]$ contains all the probability parameters for each transformations. In our setup, $\Lambda$ and $P$ can be further separated into $a=[\Lambda_a;P_a]$ and $d=[\Lambda_d;P_d]$ for spatial and domain transformations respectively.

\subsection{Cross Reconstruction}\label{section:CrossRec}

According to experimental findings, the MI loss indeed forces the two representations to have minimal shared information. However, the minimization of MI between the anatomical and domain features cannot necessarily make both features contain the respective information. The network goes into local optimums frequently, where the domain features are kept constant, and all the information is stored in the anatomical features. Because there is no information in the domain features, the MI between two representations is thus approaching zero. However, this is not our original intention. As a result, cross reconstruction strategy is introduced to tackle this problem.
The cross reconstruction loss will punish the behavior of summarizing all the information into one representation. Thus, it can force each encoder to extract informative features accordingly and prevent the whole network from going into the local optimums. 
\section{Experiments}

\subsection{Implementation Details}\label{section:implementation}
The training dataset consists of $2107$ carotid US images of one adult acquired using Siemens Juniper US Machine (ACUSON Juniper, SIEMENS AG, Germany) with a system-predefined "Carotid" acquisition parameter. The test dataset consists of (1) ValS: $200$ carotid US images which are left out from the training dataset, (2) TS1: $538$ carotid US images of $15$ adults from Ultrasonix device, (3) TS2: $433$ US images of $2$ adults and one child from Toshiba device, and (4) TS3: $540$ US images of $6$ adults from Cephasonics device (Cephasonics, California, USA). TS1 and TS2 are from a public database of carotid artery~\cite{vriha2013novel}. Notably, due to the absence of annotations, the publicly accessed images were also annotated by ourselves under the supervision of US experts. 
The acquisition was performed within the Institutional Review Board Approval by the Ethical Commission of the Technical University of Munich (reference number 244/19 S).
All the images are resized to $256\times256$ for training and testing.

We use Adam optimizer with a learning rate of $1\times10^{-4}$ to optimize all the parameters. The training is carried out on a single GPU (Nvidia TITAN Xp) with 12GB memory.

\subsection{Performance Comparison on Unseen Datasets}
In this section, we compare the performance of the proposed MI-SegNet with other state-of-art segmentation networks. All the networks are trained on the same dataset described in Section~\ref{section:implementation} with $200$ episodes. 

\subsubsection{Without Adaptation}\label{section:WithoutFineTuning}: The trained models are then tested directly on 4 different datasets described in Section~\ref{section:implementation} without further training or adaptation on the unseen domains. The dice score (DSC) is applied as the evaluation metrics. The results are shown in Table~\ref{tab1}.

\begin{figure}[h]
\includegraphics[width=\textwidth]{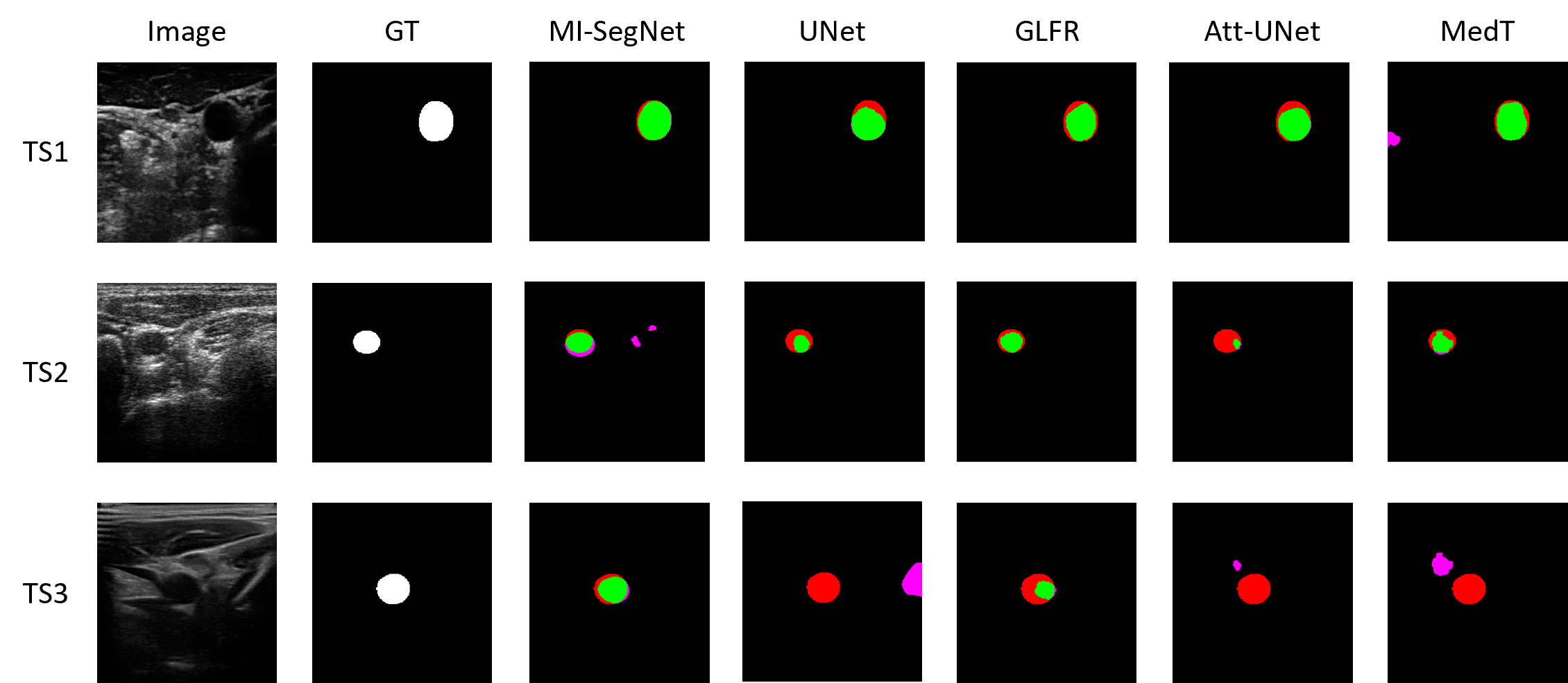}
\caption{Visual comparison between MI-SegNet and other segmentation networks on US carotid artery datasets without adaptation. For each row, we show the input US image, the ground truth (GT), and the output of each network. Red, pink and green regions represent the false negative, false positive and true positive, respectively.
}\label{Fig:VisualComparison}
\end{figure}

\begin{table}
\caption{Performance comparison of the proposed MI-SegNet with different segmentation networks on the US carotid artery datasets without adaptation.}\label{tab1}
  \begin{tabular}{L{0.30\textwidth} C{0.17\textwidth}| C{0.17\textwidth}| C{0.17\textwidth}| C{0.17\textwidth}}
    \toprule
    \multirow{2}{*}{Method}&\multicolumn{4}{c}{DSC}\\
    \cmidrule{2-5}
     & ValS & TS1 & TS2 & TS3\\
    \midrule
    UNet~\cite{ronneberger2015u}            & 0.920$\pm$0.080 & 0.742$\pm$0.283 & 0.572$\pm$0.388 & 0.529$\pm$0.378\\
    GLFR~\cite{song2022global}              & 0.927$\pm$0.045 & 0.790$\pm$0.175 & 0.676$\pm$0.272 & 0.536$\pm$0.347\\
    Att-UNet~\cite{schlemper2019attention}  & \textbf{0.932$\pm$0.046} & 0.687$\pm$0.254 & 0.602$\pm$0.309 & 0.438$\pm$0.359\\
    MedT~\cite{valanarasu2021medical}       & 0.875$\pm$0.056 & 0.674$\pm$0.178 & 0.583$\pm$0.303 & 0.285$\pm$0.291\\
    MI-SegNet w/o $\mathcal{L}_{MI}$        & 0.928$\pm$0.057 & 0.768$\pm$0.217 & 0.627$\pm$0.346 & 0.620$\pm$0.344 \\
    MI-SegNet w/o cross rec.                & 0.921$\pm$0.050 & 0.790$\pm$0.227 & 0.662$\pm$0.309 & 0.599$\pm$0.344 \\
    MI-SegNet                               & 0.928$\pm$0.046 & \textbf{0.821$\pm$0.146} & \textbf{0.725$\pm$0.215} & \textbf{0.744$\pm$0.251}\\
    \bottomrule
  \end{tabular}
\end{table}
Compared to the performance on ValS, all networks demonstrate a performance degradation on unseen datasets (TS1, TS2, and TS3). In order to validate the effectiveness of the MI loss as well as the cross reconstruction design, two ablation networks (MI-SegNet w/o $\mathcal{L}_{MI}$ and MI-SegNet w/o cross rec.) are introduced here for comparison.
The visual comparisons are shown in Fig.~\ref{Fig:VisualComparison}.
The results on TS1 are the best among all three unseen datasets while the scores on TS3 are the worst for most networks, which indicates that the domain similarity between the source and target domain decreases accordingly from TS1 to TS3.
The MI-SegNet performs the best among others on all three unseen datasets, which showcases the high generalization ability of the proposed framework. 

\subsubsection{After Adaptation}\label{section:afterFineTuning}: Although the proposed network achieves the best scores when applied directly to unseen domains, performance decay still occurs. Using it directly to unseen dataset with degraded performance is not practical. As a result, adaptation on the target domain is needed. The trained models in Section~\ref{section:WithoutFineTuning} are further trained with $5\%$ data of each unseen test dataset. The adapted models are then tested on the rest $95\%$ of each dataset. Notably, for the MI-SegNet only the anatomical encoder and segmentor are involved in this adaptation process, which means the network is updated solely based on $\mathcal{L}_{seg}$.

\begin{table}
\caption{Performance comparison of the proposed MI-SegNet with different segmentation networks after adaptation when 5\% data of each test dataset is used for adaptation.}\label{tab2}
  \begin{tabular}{L{0.24\textwidth} C{0.24\textwidth}| C{0.24\textwidth}| C{0.24\textwidth}}
    \toprule
    \multirow{2}{*}{Method}&\multicolumn{3}{c}{DSC}\\
    \cmidrule{2-4}
     & TS1 & TS2 & TS3\\
    \midrule
    UNet~\cite{ronneberger2015u}            & $0.890\pm0.183$ & $0.707\pm0.328$ & $0.862\pm0.139$\\
    GLFR~\cite{song2022global}              & $0.915\pm0.154$ & $0.875\pm0.099$ & $0.907\pm0.049$\\
    Att-UNet~\cite{schlemper2019attention}  & $0.916\pm0.117$ & $0.876\pm0.145$ & $0.893\pm0.087$\\
    MedT~\cite{valanarasu2021medical}       & $0.870\pm0.118$ & $0.837\pm0.137$ & $0.795\pm0.170$\\
    MI-SegNet                               & $\mathbf{0.919\pm0.095}$ & $\mathbf{0.881\pm0.111}$ & $\mathbf{0.916\pm0.061}$\\
    \bottomrule
  \end{tabular}
\end{table}

The intention of this experiment is to validate whether the proposed network can serve as a good pre-trained model for the downstream task. A well-trained pre-trained model, which can achieve good results when only a limited amount of annotations is provided, has the potential to release the burden of manual labeling and adapts to different domains with few annotations.
Table~\ref{tab2} shows that the MI-SegNet performs the best on all test datasets.
However, the difference is not that significant as in Table~\ref{tab1} when no data is provided for the target domain. This is partially due to the fact that carotid artery is a relatively easy anatomy for segmentation. It is observed that when more data (10\%) is involved in the adaptation process GLFR and Att-UNet tend to outperform the others and it can be therefore expected when the data size further increases all the networks will perform equally well on each test set.

\section{Discussion and Conclusion}
In this paper, we discuss the particular importance of domain adaptation for US images. Due to the low speed of sound compared to light and X-ray, the complexity of US imaging and its dependency on many parameters are more remarkable than optical imaging, X-ray, and CT. Therefore, the performance decay caused by the domain shift is a prevalent issue when applying DNNs in US images. To address this problem, a MI-based disentanglement method is applied to increase the generalization ability of the segmentation networks for US image segmentation.
The ultimate goal of increasing the generalizability of the segmentation network is to apply the network to different unseen domains directly without any adaptation process. However, from the authors' point of view, training a good pre-trained model that can be adapted to an unseen dataset with minimal annotated data is still meaningful. As demonstrated in Section~\ref{section:afterFineTuning}, the proposed model also shows the best performance in the downstream adaptation tasks. Currently, only the conventional image transformation methods are involved. In the future work, more realistic and US specific image transformations could be implemented to strengthen the feature disentanglement.


\bibliographystyle{splncs04}
\bibliography{references}

\end{document}